\documentclass[twocolumn,pre,aps,showpacs,amsmath,amssymb,superscriptaddress]{revtex4}

\usepackage{hyperref}
\usepackage{amsmath,amssymb}
\usepackage{amsfonts,amsthm}
\usepackage{graphics}
\usepackage{graphicx}
\usepackage{dcolumn}
\usepackage{color}
\usepackage{bm}

\usepackage[normalem]{ulem}

\begin{document}


\title{Phase-bistability between anticipated and delayed synchronization in neuronal populations}

\author{Julio Nunes Machado}
\affiliation{Instituto de F\'{\i}sica, Universidade Federal de Alagoas, Macei\'{o}, Alagoas 57072-970 Brazil.}

\author{Fernanda Selingardi Matias}
\thanks{fernanda@fis.ufal.br}
\affiliation{Instituto de F\'{\i}sica, Universidade Federal de Alagoas, Macei\'{o}, Alagoas 57072-970 Brazil.}

\begin{abstract}

Two dynamical systems unidirectionally coupled in a sender-receiver configuration can synchronize with a nonzero phase-lag. In particular, the system
can exhibit anticipated synchronization (AS), which is characterized by a negative phase-lag, if the receiver (R) also receives a delayed negative self-feedback. 
Recently, AS was shown to occur between cortical-like neuronal populations in which the self-feedback is mediated by inhibitory synapses.
In this biologically plausible scenario, a transition from the usual delayed synchronization (DS, with positive phase-lag) to AS
can be mediated by the inhibitory conductances in the receiver population.
Here we show that depending on the relation between excitatory and inhibitory synaptic conductances the system can also exhibit 
phase-bistability between anticipated and delayed synchronization.
Furthermore, we show that the amount of noise at the receiver and the synaptic conductances 
can mediate the transition from stable phase-locking to a bistable regime and eventually to a phase-drift (PD).
We suggest that our spiking neuronal populations model could be potentially useful to study phase-bistability in cortical regions related to bistable perception.

\end{abstract}
\maketitle

%
%

\section{Introduction}

Multistable perception is the brain's ability to alternate between two or more perceptual states that occur when sensory information is ambiguous~\cite{Leopold99}.
It has been hypothesized that the perception of two different features of the same visual cue such as in the Necker cube~\cite{Necker1832}
(see Fig.~\ref{fig:motif}(a), which can be interpreted to have either the upper-left or the lower-right square as its front side)
should be related to the nonlinear interaction between brain rhythms~\cite{Battaglia12}. 
This means that an almost fixed anatomical connectivity should allow flexible changes from one functional connectivity 
pattern to another on timescales relevant to behavior. 
Recently it has been reported that bistable phase-differences in magnetoencephalography (MEG) recordings appear when participants listening to bistable speech sequences that could be perceived as two distinct word sequences repeated over time~\cite{Kosem16}.
This result suggests that phase-bistability in cortical regions could be related to bistable perception.

Here we propose a model of two neuronal populations that can synchronize with a bistable phase-lag.
Typically, when two dynamical systems are unidirectionally coupled they can synchronize with a positive phase-lag in which the sender is also the leader. 
This regime is called delayed synchronization (DS).
However, it has been shown that in a sender-receiver configuration if the receiver is subjected to a delayed self-feedback the system can synchronize with a negative phase-lag~\cite{Voss00}. 
This counterintuitive situation indicates that the receiver leads the sender and it is called anticipated synchronization (AS). 
Formally, Voss has shown that if a system is described by the following equations:
\begin{eqnarray}
\label{eq:voss}
\dot{\bf {S}} & = & {\bf f}({\bf S}(t)), \\
\dot{\bf {R}} & = & {\bf f}({\bf R}(t)) + {\bf K}[{\bf S}(t)-{\bf R}(t-t_d)], \nonumber 
\end{eqnarray}
with arbitrary continuous ${\bf f}$ and coupling matrix ${\bf K}$, it may has a stable solution ${\bf R}(t)={\bf S}(t-t_d)$, which characterizes AS i.e. the receiver predicts the sender. 

In the last two decades AS has been verified in both theoretical~\cite{Voss00,Voss01b,Voss01a,Ciszak03,Masoller01,HernandezGarcia02,Sausedo14,Voss16,Voss16Negative,Voss18} and
experimental~\cite{Sivaprakasam01,Ciszak09,Tang03} studies. AS can also occur if the delayed self-feedback is replaced by different parameter mismatches at the receiver~\cite{Kostur05,Pyragiene13,Simonov14}, 
including inhibitory loops mediated by chemical synapses~\cite{Matias11,Matias15,Matias16,Matias17,Mirasso17,Pinto19}.
Especially it has been shown that a faster internal dynamics of the receiver could promote AS between unidirectionally coupled oscillators~\cite{Hayashi16,Dima18,Pinto19,DallaPorta19}.

Recently, AS has been verified between unidirectionally coupled cortical-like populations~\cite{Matias14,DallaPorta19}. 
The neuronal population model exhibits a smooth transition from DS to AS that can be mediated by synaptic conductances as well as by external noise.
The model could explain electrophysiological results in non-human primates showing that
unidirectional Granger causality can be accompanied by both positive or negative phase difference between cortical areas~\cite{Matias14,Montani15,Brovelli04,Salazar12}.
Furthermore, AS in the alpha band has been recently reported between synchronized electrodes in human EEG~\cite{Carlos20} as well as in human behavior~\cite{Washburn19,Roman19}.

Here we show that a simple biologically plausible motif with two unidirectionally coupled neuronal populations 
can exhibit phase-bistability between delayed and anticipated synchronization regimes.
Especially our populations present a bimodal distribution of phase-differences with a positive (DS) and a negative (AS) peak. 
In Sec.~\ref{model} we describe our motif as well as neuronal and synaptic models.
In Sec.~\ref{results}, we report our results, showing that the phase-bistability can be mediated by synaptic couplings and noise.
Concluding remarks and a brief discussion of the significance of our findings for neuroscience are presented in Sec.~\ref{conclusions}.

\section{\label{model}Neuronal populations model}


\begin{figure}[!ht]
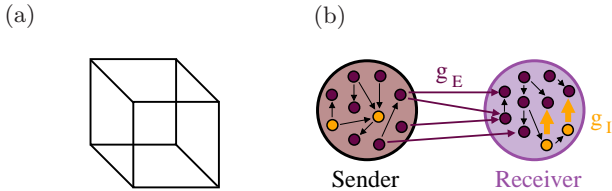
%
 \begin{minipage}{4cm}
  \begin{flushleft}(a)%
\end{flushleft}%
\includegraphics[width=0.44\columnwidth,clip]{Nunes01a}
\end{minipage}
\begin{minipage}{4cm}
\begin{flushleft}(b)%
\end{flushleft}%
\includegraphics[width=0.99\columnwidth,clip]{Nunes01b}
\end{minipage}
\caption{\label{fig:motif} (Color online) (a) Necker cube~\cite{Necker1832} as an example of bistable perception. 
Both the upper-left and the lower-right square can be the front side. (b) Schematic representation of two cortical areas coupled in a sender-receiver configuration. 
The inhibitory feedback is controlled by the synaptic conductance $g_I$ at the receiver population.
  }
\end{figure}%

Our neuronal motif is composed of two unidirectionally coupled cortical-like 
neuronal populations: a sender (S) and a receiver (R),
see Fig.~\ref{fig:motif}(b). Each one is composed of 400 excitatory and 100 inhibitory
neurons~\cite{Matias14} described by the Izhikevich
model \cite{Izhikevich03}:
\begin{eqnarray}
  \frac{dv}{dt} &=& 0.04v^2+5v+140-u +\sum_x I_{x}, \label{dv/dt}\\
  \frac{du}{dt} &=& a(bv-u). \label{du/dt}
\end{eqnarray}
In Eqs.~\ref{dv/dt} and~\ref{du/dt} $v$ is the membrane potential and
$u$ the recovery variable which accounts for activation
of K$^+$ and inactivation of Na$^+$ ionic currents.  $I_x$ are the synaptic currents provided by
the interaction with other neurons and external inputs.  If
$v\geq30$~mV, $v$ is reset to $c$ and $u$ to $u+d$.  To account for
the natural heterogeneity of neuronal populations, which can exhibit a variety of neuronal dynamics (regular spiking, 
bursting, chatering, fast spiking, etc.~\cite{Izhikevich04a}), 
the dimensionless parameters are randomly sampled as follows: $(a,b)=(0.02,0.2)$ and
$(c,d)=(-65,8)+(15,-6)\sigma^2$ for excitatory neurons and $(a,b)=(0.02,0.25)+(0.08,-0.05)\sigma$ and
$(c,d)=(-65,2)$ for inhibitory neurons, where $\sigma$ is a
random variable uniformly distributed on the interval [0,1]~\cite{Izhikevich03,Izhikevich04a}. 
Equations were integrated with the Euler method and a time step of $0.05$~ms.

The connections between neurons in each population are assumed to be
fast unidirectional excitatory and inhibitory chemical synapses
mediated by AMPA and GABA$_\text{A}$. The synaptic currents are given
by:
\begin{equation}
\label{Ix}
I_{x} = g_{x}r_{x}(v-V_x),
\end{equation}
where $x=E,I$ (excitatory and inhibitory mediated by AMPA and
GABA$_\text{A}$, respectively), $V_E=0$~mV, $V_I=-65$~mV, $g_{x}$ is the maximal
synaptic conductance and $r_{x}$
is the fraction of bound synaptic receptors whose dynamics is given
by:
\begin{equation}
\label{drdt}
  \tau_x\frac{dr_{x}}{dt}=-r_{x} + D \sum_k \delta(t-t_k),\\
\end{equation}
where the summation over $k$ stands for pre-synaptic spikes at times
$t_k$. D is
taken, without loss of generality, equal to $0.05$. The time decays are $\tau_{E}=5.26$~ms $\tau_{I}=5.6$~ms.  Each
neuron is subject to an independent noisy spike train described by a
Poisson distribution with rate $R$. The input mimics excitatory synapses 
(with conductances $g_{P}$) from $n$ pre-synaptic neurons external to the population,
each one spiking with a Poisson rate $R/n$ which, together with a constant external current $I_c$,  
determine the main frequency of mean membrane potential of each population. Unless otherwise stated, we have employed $R=2400$~Hz and $I_c=0$. 
We have fixed the Poissonian synaptic conductance at the sender population $g_{P}^{S}=0.5$~nS and varied $g_{P}$ only at the receiver population.  

Connectivity within each
population randomly targets 10\% of the neurons, with excitatory
conductances set at $g^{S}_{E} =g^{R}_{E} = 0.5$~nS. Inhibitory conductances are fixed at the sender population $g^{S}_{I} = 4.0$~nS 
and $g_I$ at the receiver population is varied throughout the study (see Fig.~\ref{fig:motif}).
Each neuron at the R population receives 
20 fast synapses (with conductance $g_{E}$) from random excitatory
neurons of the S population.
The bistability studied in this paper only happens when the synaptic conductances $g_{E}$, $g_{I}$, and $g_{P}$ have comparable values.
The phase-locking regimes presented by the model for $g_{E} \geq 0.5$~nS and $g_{I} \geq 1.0$~nS have been studied in Ref.~\cite{Matias14}.

\subsection*{Characterizing time delay in the model}
The mean membrane potential $V_x$ ($x=$S, R) is calculated as the total sum of the membrane potential 
$v$ of each neuron in the $x$ population in a given time $t$, divided by the total number of neurons in that population.
Fig.~\ref{fig:DSAS}(a) and (b) show examples of $V_S$ and  $V_R$ in different regimes. 
Since $V_x$, which we assume as a crude approximation of the measured local field potential (LFP), is
noisy, we average within a sliding window of width 5-8~ms to obtain a smoothened signal, from which we can extract 
the peak times $\{t^{x}_{i}\}$ (where $i$ indexes the peak). The
period of a given population in each cycle is thus $T^{x}_i \equiv t^{x}_{i+1}-t^{x}_{i}$. 
For a sufficiently long time series, we compute the mean period
$T_x$ and its variance.  

In a similar way we can define the time delay between the sender and the receiver populations in each cycle
$\tau_i=t^{R}_{i}-t^{S}_{i}$ (Fig.~\ref{fig:DSAS}(a) and (b)). Then, if $\tau_i$ obeys a unimodal distribution,  we calculate $\tau$ as the mean
value of $\tau_i$ and $\sigma_{\tau}$ as its variance.  
If $T_S  \approx  T_R $
and $\tau$ is independent of the initial conditions, the populations exhibit
oscillatory synchronization with a phase-locking regime. In all those calculations we discard the
transient time.

\section{Results}
\label{results}

\subsection{Phase-locking regimes: delayed and anticipated synzhronization}
\label{DSAS}

\begin{figure}
\centerline{\includegraphics[width=0.86\columnwidth,clip]{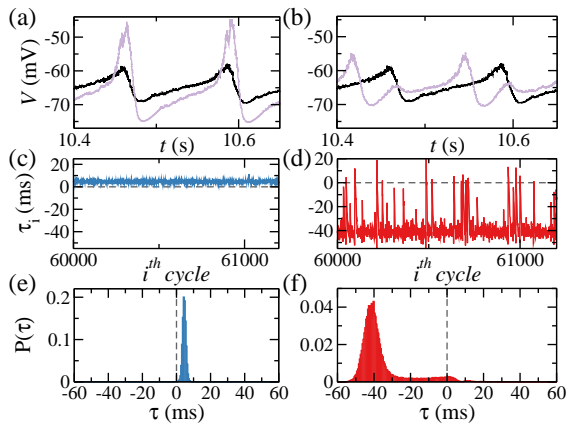}}
\caption{\label{fig:DSAS} 
Characterizing the phase-locking regimes. Left column: delayed synchronization (DS, $g_E=0.8$~nS and $g_I=0.02$~nS) in which the sender population leads 
the receiver on average with a positive mean time delay $\tau=4.5$~ms. 
Right column: anticipated synchronization (AS, $g_E=0.5$~nS and $g_I=0.8$~nS) in which the receiver population leads de sender on average with a negative mean time delay $\tau=-35.8$~ms. 
(a) and (b) Mean membrane potential of each neuronal population (the sender in black and the receiver in light purple).
(c) and (d) Time delay $\tau_i$ in each cycle.
(e) and (f) Histograms of time delays per cycle.
}
\end{figure}

In order to mimic the oscillatory activity in cortical regions 
we simulated the sender population in such a way that the external noise 
and the internal coupling are enough to allow the mean membrane potential to oscillate with $f\simeq8$~Hz (equivalent to $T^{S}_i\simeq 125$~ms).
Depending on the internal parameters of the receiver population, the sender-receiver coupling $g_E$ can synchronize the activity of both areas or not. 
The phase-locked regimes present non-zero phase-lags. 
In Fig.~\ref{fig:DSAS}(a) and (b) we show simulated time series of the S and R population. 
Fig.~\ref{fig:DSAS}(c) and (d) show the time delay $\tau_i$ in the $i-th$ period as a function of the period index $i$ 
and Fig.~\ref{fig:DSAS}(e) and (f) display their probability densities.

The phase-lockings can be characterized by the mean time delay $\tau$ and its standard deviation.
For sufficiently large $g_E$ the mean time delay $\tau$ is positive which indicates that the sender population leads the receiver. 
This is the usual delayed synchronization regime (DS).
Left panels of Fig.~\ref{fig:DSAS} show an example of DS for $g_E=0.8$~nS and $g_I=0.02$~nS. 
For this set of parameters, the peak at the receiver population occurs on average $\approx 4.5$~ms after the peak of the sender, 
which is close to the magnitude of the synaptic time scales (as mentioned in Sec.~\ref{model}: $\tau_{E}=5.26$~ms)

For larger values of inhibition $g_I$, the receiver population leads the sender on average, which is characterized by a negative mean time delay: $\tau<0$. 
This non-intuitive situation is the so-called anticipated synchronization~\cite{Voss00,Matias14}. 
The right panels in Fig.~\ref{fig:DSAS} exhibit an example of AS with $g_E=0.5$~nS and $g_I=0.8$~nS.
For this set of parameters, the system exhibits $\tau\simeq-35.8$~ms. Such value could not be inferred from any model parameter as, for example, the synaptic time scales. 
This means that the anticipation time is a result of the nonlinear dynamics of the system.
Furthermore, the anticipation does not occur every cycle, but the distribution of $\tau_i$ (see  Fig.~\ref{fig:DSAS}(f)) clearly 
shows that in the AS regime the majority of the peaks happens in the receiver-sender order. 

\subsection{Bistability between DS and AS}
\label{Bistability}

\begin{figure}[t]
\centerline{\includegraphics[width=0.78\columnwidth,clip]{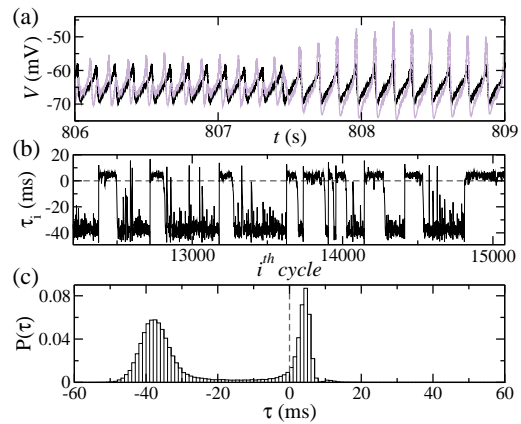}}
\caption{\label{fig:bistable} 
Characterizing the phase-bistability (excitatory and inhibitory conductances respectively $g_E=0.6$~nS and $g_I=0.4$~nS).  
(a) Mean membrane potential of S and R populations during a transition from AS to DS. 
(b) Time delay per cycle which alternates from positive to negative time delays every few cycles in a non deterministic way. 
(c) Bimodal probability distribution of time delays characterizing the bistable phase.
}
\end{figure}

A system that spends the vast majority of
time near two well-separated regions exhibits a stationary density with two sharp peaks. 
This phenomenon characterizes bistability.
The Kramers problem is a standard example of a bistable system,
which consists of diffusion in a double-well potential in the presence of noise.
The distribution of the time spent in each region before a transition
depends on the amount of noise.

Here we show that depending on the relation between excitatory and inhibitory conductances the system can present a bistable regime between a DS and an AS regime (see Fig.~\ref{fig:bistable}). 
The time delay $\tau_i$ is positive for a few cycles, with a well-defined mean value and standard deviation, which is similar to a DS regime for a certain amount of time. 
Then, the system randomly switches for different dynamics in which $\tau_i$ is negative during a few other cycles. By analyzing the system only for these few periods
of oscillation, one could wrongly characterize the system as in an AS regime. But, then, suddenly again, the system can jump back to the first attractor close to DS.
Therefore, the histogram of time delays between the two populations is a bigaussian with one positive and one negative peak as shown in Fig.~\ref{fig:bistable}(c).
In this regime, the system cannot be simply characterized by the mean time delay $\tau$.

If the system remains close to the DS-attractor (AS-attractor) for more than two cycles we define this as a DS-event (AS-event). 
The DS-events are represented by the upper states in Fig.~\ref{fig:bistable}(b) whereas AS-events are the lower states in the figure.
We define the event size as the number of cycles in which the system stays close to one of the two regions. 
Since the mean period of oscillation of the sender is $125$~ms we can use the size of the event as a proxy for the temporal dynamics of the bistable regime. 
For example, an event that lasts for 8 cycles would have $\approx1$~s duration.
This would be especially useful if one needs to compare the temporal dynamics of the model with behavioral data.

\begin{figure}[h]
\centerline{\includegraphics[width=0.8\columnwidth,clip]{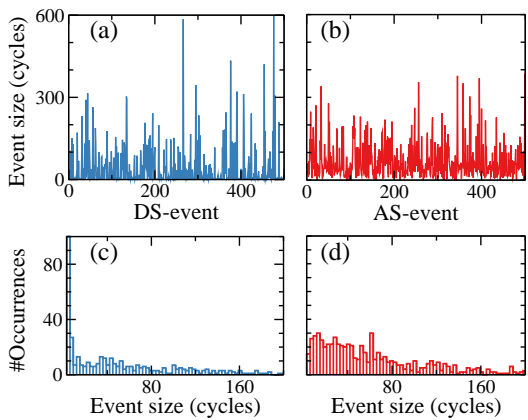}}
\caption{\label{fig:bistabletime} 
{\bf Temporal dynamics of the bistability.}
 Same parameters as in Fig.~\ref{fig:bistable}: $g_E=0.6$~nS and $g_I=0.4$~nS.  
(a) Size of each DS-event in number of cycles (or periods of the sender) that the system remains close to the DS atrractor.
These values could be converted to time since each period takes $\approx125$~ms long.
(b) Size of the AS-events.   
(c) Distribution of the size events close to DS and
(d) close to AS.
}
\end{figure}

In Fig.~\ref{fig:bistabletime} we show the size of the events and their histograms.
The distribution of the number of occurrences of a specific size is different for DS-events and AS-events. 
The probability to find very small events (up to 9 cycles) or very large events (larger than 300 cycles) is larger for DS than for AS-events. 
On the other hand, events of intermediate sizes are more probable close to the AS region. 
The distributions are qualitatively comparable to temporal dynamics of binocular rivalry during fMRI~\cite{Lumer98}.
However, investigating in more detail the statistical properties of these distributions, to compare with cognitive data would require a significative computational effort to simulate even longer time series which is out of the scope of this study.

\subsection{Phase-drift regime}

\begin{figure}
\centerline{\includegraphics[width=0.8\columnwidth,clip]{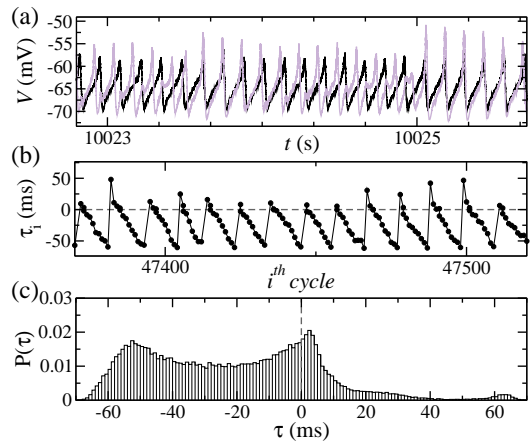}}
\caption{\label{fig:PD} 
Example of a phase-drift regime in which the receiver is faster than the sender ($g_E=0.3$~nS and $g_I=0.4$~nS). (a) Mean membrane potential of S and R populations. 
(b) Time delay per cycle. 
(c) Distribution of time delays characterizing the phase-drift.
}
\end{figure}

For small values of the sender-receiver coupling $g_E$, the system can also exhibit a phase-drift (PD) regime in which the receiver is faster than the sender ($T_S  > T_R $).
Fig.~\ref{fig:PD} shows an example of such a regime:
the time delay $\tau_i$ changes every cycle in a quasi-periodic configuration.
The histogram of $\tau_i$ for $g_E=0.3$~nS seems flattened. 
In the limit of $g_E=0$, which characterizes the totally uncoupled situation,
every $\tau_i$ is equiprobable. 
As in the bistable regime, in the PD, we do not use the mean time delay $\tau$ to characterize the regime.

\begin{figure}[h]
\centering
\begin{minipage}{0.49\linewidth}
\begin{flushleft}(a)%
\end{flushleft}%
\centering
\includegraphics[width=0.98\columnwidth,clip]{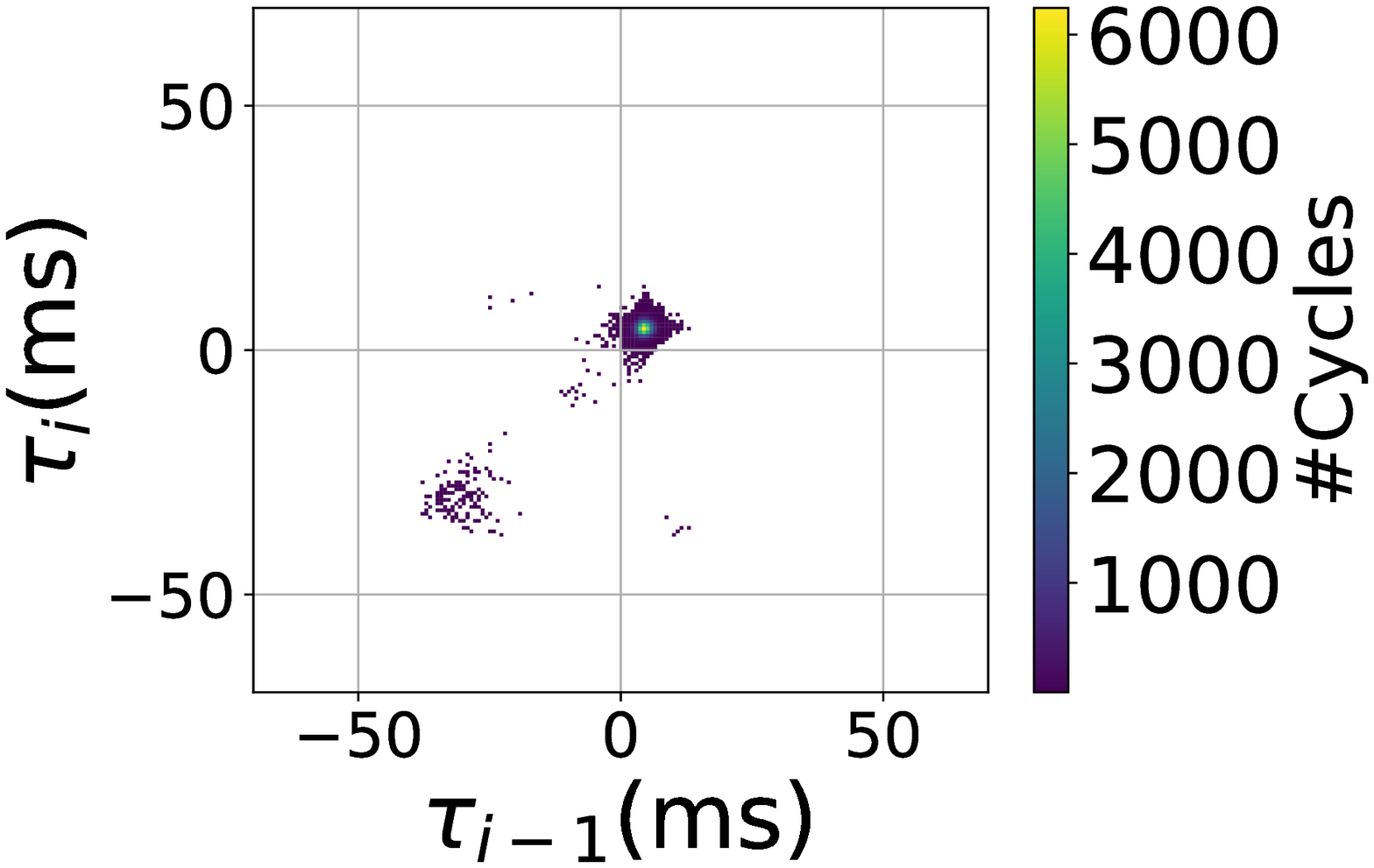}
\end{minipage}
\begin{minipage}{0.49\linewidth}

\begin{flushleft}(b)%
\end{flushleft}%
\centering
\includegraphics[width=0.98\columnwidth,clip]{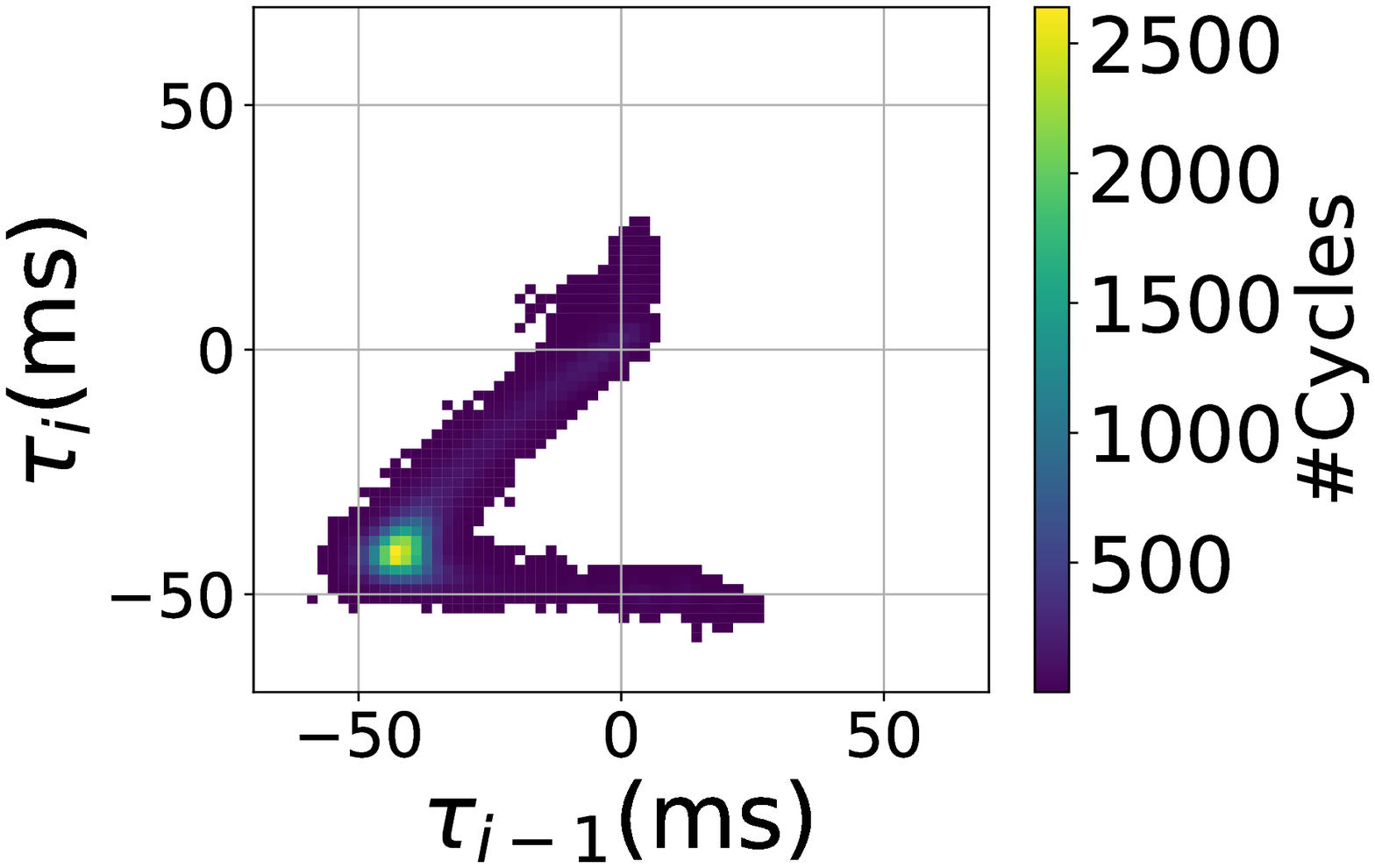}
\end{minipage}
\begin{minipage}{0.49\linewidth}

\begin{flushleft}(c)%
\end{flushleft}%
\centering
\includegraphics[width=0.98\columnwidth,clip]{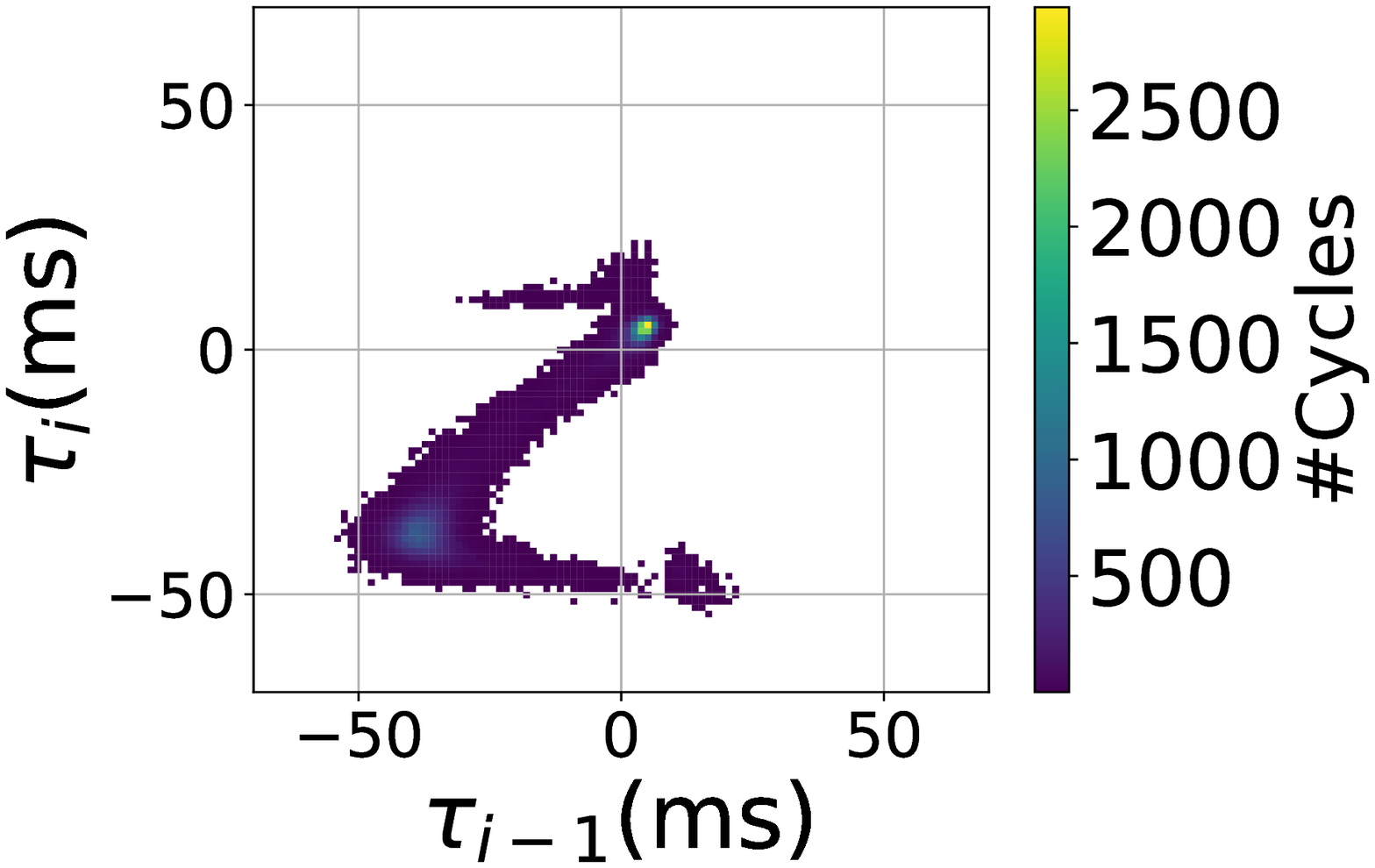}
\end{minipage}
\begin{minipage}{0.49\linewidth}

\begin{flushleft}(d)%
\end{flushleft}%
\centering
\includegraphics[width=0.98\columnwidth,clip]{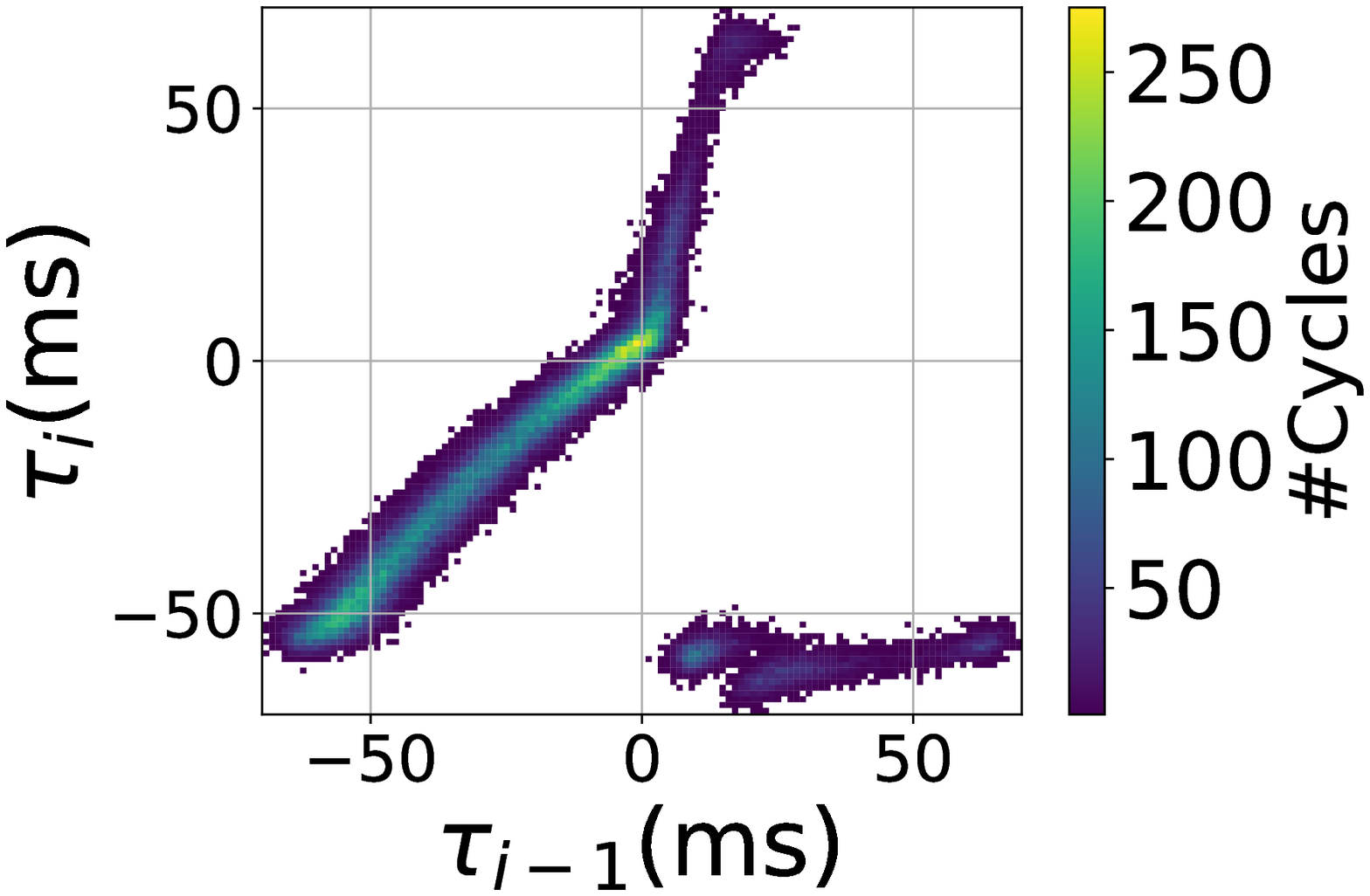}
\end{minipage}
\caption{
{\bf  The return map of the time delays $\tau_i$ versus $\tau_{(i-1)}$ in a heatmap for each regime:}
(a) delayed synchronization, (b) anticipated synchronization, (c) phase-bistability and (d) phase-drift. 
Same parameters as in previous examples of Figs.\ref{fig:DSAS}, \ref{fig:bistable} and \ref{fig:PD}.
}
\label{fig:heatmap}
\end{figure}

\begin{figure}[ht]
\centering
\includegraphics[width=0.9\columnwidth,clip]{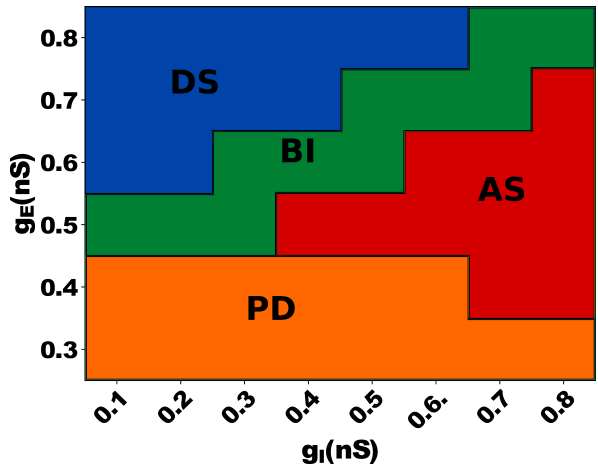}
\caption{\label{fig:diagram} 
{\bf Two-dimensional projections of the phase diagram of our model: DS (blue), AS (red), bistability (BI, green), PD (orange).}
The effect of the relation between excitation and inhibition at the receiver is shown by $g_E$ \textit{versus} $g_I$. 
For this set of parameters, the transition from DS to AS only occurs via bistability. For small enough $g_E$ the system eventually reaches a phase-drift.
}
\end{figure}

In Fig.~\ref{fig:heatmap} we display a heatmap of the return map $\tau_i$ \textit{versus} $\tau_{i-1}$ which is useful to illustrate the different dynamical regimes.
Several features in those curves are important to mention.
Due to the noise, in every regime, there is a probability to find both positive and negative time delays per cycle $\tau_i$. 
However, in the DS regime, the probability density is clearly larger at the first quadrant, whereas in the AS the system remains for longer times in the third quadrant.
Therefore, the bistable regime presents two denser regions: one in the first and the other in the third quadrant.
In the PD regime there are continuous regions of the return map that are almost equally accessed by the system (see Fig.~\ref{fig:heatmap}(d)). 
This reflects the fact that the time delay varies from positive to negative in small steps during a few cycles (see Fig.~\ref{fig:PD}(b)).


\subsection{\label{sec:parameters} Scanning parameter space}

The dependence of the system's dynamics on the synaptic conductances $g_E$ and $g_I$ is shown in
Fig.~\ref{fig:diagram}. As one could expect, for small enough sender-receiver coupling there is no phase-locked regime. 
As we decrease the $g_E$ the system eventually reaches the PD.
It is worth emphasizing that, similarly 
to previous results on AS~\cite{Matias14,DallaPorta19}, 
the transition from DS to AS can be mediated by both internal inhibition and external noise at the receiver.
However, differently from the previous studies~\cite{Matias14,DallaPorta19}, here, the DS-AS transition
does not occur via zero-lag synchronization,
but through a bistable regime.

\begin{figure}[!h]
\centering
\includegraphics[width=0.9\columnwidth,clip]{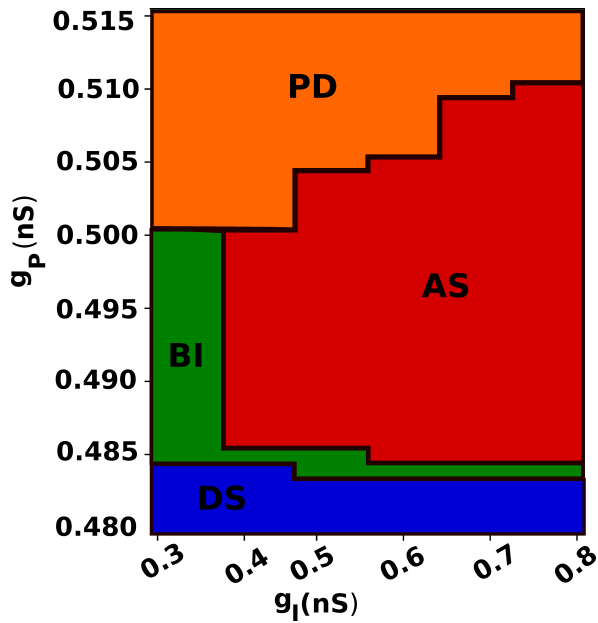}
\caption{\label{fig:gpgi} 
{\bf  The effect of noise at de receiver population is summarized at the ($g_I$, $g_P$) projection of parameter space: }
DS (blue), AS (red), bistability (BI, green), PD (orange). The sender-receiver coupling is $g_E=0.5$~nS. 
}
\end{figure}

To better understand the effects of external noise in the dynamical regimes, we vary the synaptic conductance $g_P$
of the Poissonian input received by each neuron of the receiver population. This external input mimics synaptic currents received from  other 
cortical regiones that are not included in our simple two-population model (see Sec.~\ref{model} for more details).
Fig.~\ref{fig:gpgi} displays a two dimensional ($g_I$,$g_P$) projection of parameter space.
The effect of increasing the noise at the receiver $g_P$ is similar to the effect of decreasing the sender-receiver coupling $g_E$.
In the PD regime, the time delay in each cycle varies almost continuously, 
whereas in the bistable regime $\tau_i$ suddenly changes from positive to negative values in a non-predictable way.

\begin{figure}[h]
\centerline{\includegraphics[width=0.999\columnwidth,clip]{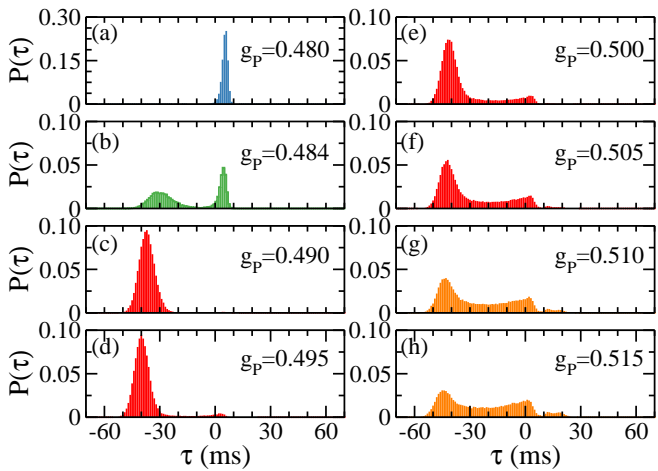}}
\caption{\label{fig:hist} 
{\bf The effect of external input in the dynamical regimes.}
Distribution os time delays  $\tau_i$ between S and R population in each cycle for different values of external noise $g_P$. 
For $g_E=0.5$~nS and $g_I=0.6$~nS the system exhibits AS if sender and receiver are subjected to the same amount of noise ($g_P=g_{P}^{S}=0.5$~nS). 
Increasing the noise at the receiver ($g_P\geqslant g_{P}^{S}$), the system undergoes a transition from AS to PD. 
Decreasing the noise at R ($g_P \leq g_{P}^{S}$) the AS gives rise to the bistable regime and eventually to DS.
}
\end{figure}

Fig.~\ref{fig:hist} shows illustrative examples of the time delay distribution as we change the external Poissonian input at the receiver population $g_P$ 
(along the vertical line  $g_I=0.6$~nS in Fig.~\ref{fig:gpgi}. ). 
By choosing conductances in such a way that the system presents AS when the amount of noise in both populations are the same ($g_P=g_{P}^{S}=0.5$~nS, $g_E=0.5$~nS and $g_I=0.6$~nS ), 
as we increase the noise at the R population ($g_P>0.5$~nS) the system goes to a PD regime. 
The AS-PD transition has been previously reported in neuronal microcircuits~\cite{Matias11,Pinto19} but not in neuronal populations.
On the other hand, by decreasing the noise at R ($g_P>0.484$~nS) the system undergoes an AS-DS transition via bistability. 

We vary the Poissonian conductance $g_P$ every $0.001$~nS and the other synaptic conductances $g_E$ and  $g_I$ every $0.1$~nS in order 
to capture all the qualitative important features of the diagrams in Fig.~\ref{fig:diagram} and ~\ref{fig:gpgi}
The conditions to define the boundaries between the regimes in Figs~\ref{fig:diagram} and ~\ref{fig:gpgi} have been determined 
by analyzing the time delay distributions as in Fig.~\ref{fig:hist}.
For the DS regime the only required condition is a positive $\tau$. 
For the AS regime, besides a negative $\tau$, the system should present an AS peak at least three times larger than the DS peak. 
If this is not the case, the system could be in a bistable regime or a phase-drift.
In the bistable regime the smallest peak should be at least 7 times higher than the probability to find intermediate values of $\tau_i$ between the peaks, otherwise it is a PD.



\section{\label{conclusions}Concluding remarks}

To summarize, we have shown that a simple but biophysically plausible model 
of two unidirectionally connected neuronal populations can present phase-bistability between anticipated and delayed synchronization.
To the best of our knowledge, this the first verification of such a regime.
Unlike previous studies on AS in neuronal models~\cite{Matias14,DallaPorta19,Matias11,Pinto19}, 
here, the transition from DS to AS does not occur via zero-lag synchronization, but through the bistable regime.
We have also shown that the interplay among the inhibitory conductance at the receiver, 
the sender-receiver coupling and the external noise determines the dynamical regime.
Moreover, for sufficiently large noise or small coupling the system can eventually reach a phase-drift regime in which the receiver population is faster than the sender.

Multi-stability of neuronal networks has been suggested as the mechanism underlying switches between
different perceptions or behaviors~\cite{Lumer98,Moreno07,Battaglia12}.
Recently, multi-stability has also been associated with different oscillatory states of brain dynamics~\cite{Freyer09,Freyer11}.
In particular, perceptual neuronal states models based on noise and adaptation have been used to qualitatively describe neurophysiological 
experiments on human visual bistable perception~\cite{Chholak20,Runnova16}. This model alternates between two different active states 
and reproduces probability distributions of dominance durations and their relation with the amount of noise. 
However, these studies were not investigating phase relations during ambiguous perception.
Therefore, we suggest that our results, using  populations of spiking neurons, could be potentially useful to study phase-bistability in cortical regions during bistable perception~\cite{Kosem16}.
In fact, our model shows fixed structural connectivity that allows flexible dynamics which could change in timescales relevant for behavior. 
As far as we know, this is the first spiking neuronal population model to present a bistable regime between two synchronized regimes with a positive and a negative phase-difference.

Our results offer a number of possibilities for further investigation. 
The DS-AS transition could possibly explain commonly reported short latency in visual systems~\cite{Orban85,Nowak95,Kerzel03,Jancke04,Puccini07,Martinez14}, 
olfactory circuits~\cite{Rospars14}, songbirds brain~\cite{Dima18} and human perception~\cite{Stepp10,Stepp17}.
Future studies can be conducted in the light of understanding the functional significance of the diversity in the phase-relations between oscillatory brain regions~\cite{Maris16}.
Furthermore, including the effects of synaptic plasticity, especially spike-timing-dependent plasticity~\cite{Abbott00,Clopath10,Matias15} 
in the bistable regime is a natural next step which we are currently pursuing.



\begin{acknowledgments}
The authors thank FAPEAL, UFAL, CNPq (grant 432429/2016-6) and CAPES (grant 88881.120309/2016-01) for finnancial support.

\end{acknowledgments}
%

\bibliography{matias}

\end{document}